\documentclass[twocolumn,amsmath,amssymb]{revtex4}
\usepackage{bm}
\usepackage{graphicx}
\newcommand{\bra}[1]{\langle #1|} 
\newcommand{\ket}[1]{|#1\rangle}
\newcommand{\braket}[2]{\langle #1|#2 \rangle}
\begin{document}
\title{Quantum Chernoff Bound metric for the XY model at finite temperature}
\author{Damian F. Abasto$^1$} \email{abasto@usc.edu}
\author{N. Tobias Jacobson$^1$} \email{ntj@usc.edu}
\author{Paolo Zanardi$^{1,2}$}
\affiliation{
$^1$ Department of Physics and Astronomy, University of Southern California, Los Angeles, CA 90089-0484
\\
$^2$ Institute for Scientific Interchange, Viale
Settimio Severo 65, I-10133 Torino, Italy
}

\begin{abstract}
We explore the finite temperature phase diagram  of the anisotropic XY spin chain using the Quantum Chernoff Bound metric on  thermal states.  The analysis of the metric elements 
allows to easily identify, in terms of different scaling with temperature, quasi-classical and quantum-critical regions. 
These results extend recent ones obtained using the Bures metric and show that different information-theoretic notions of distance can carry the same sophisticated 
information about the phase diagram of an interacting many-body system featuring quantum-critical points.
\end{abstract}
\maketitle
\section{\label{sec:level1}Introduction}
Quantum Phase Transitions (QPTs) are characterized by a dramatic change in the ground state of a many-body system prompted by a change of the parameters $\{\lambda\}$ specifying the hamiltonian $H (\{\lambda\})$
of the system. Unlike classical phase transitions, which are driven by thermal fluctuations, quantum phase transitions occur at T = 0 and are driven solely by quantum fluctuations. In particular, second order quantum phase transitions are characterized by a gap between the ground state and first excited state that vanishes at criticality \cite{QPT}.\\
\indent The analysis of these transitions has benefited from tools of Quantum Information theory. The von Neumann entropy and fidelity applied to many-body systems can identify phase transitions and reveal different scaling behaviors at different regions of the phase diagram \cite{EMB 1}-\cite{EMB 6}. More recently it has been shown that the quantum  fidelity -- a distinguishability measure between quantum states -- can identify the quantum phase transition by comparing two ground states corresponding to slightly different values of the coupling constants $\{\lambda\}$ \cite{F 1}-\cite{za-paris}. This new approach provides an alternative to the study of phase transitions using order parameters and symmetry breaking patterns, which depends on \textit{a priori} knowledge of the physics of the problem \cite{gold}. \\
\indent Another distinguishability measure for density operators was very recently introduced: the Quantum Chernoff Bound \cite{QCB}\cite{QCB2}. Suppose there are $n$ copies each of two density matrices  $\rho$ and $\sigma$, i.e., $\rho^{\otimes n}$ and $\sigma^{\otimes n}$, and one set has been given to us. We know what $\rho$ and $\sigma$ are, but we don't know which set we received. There is always a two-outcome POVM $\{E_0, E_1\}$ that measures and distinguishes one set of $n$-copy states from the other such that the probability of making a misidentification, $P_e$, is minimized. In general, this probability will depend on the number of copies. In the asymptotic limit of $n$ very large, the probability of misidentification has a particularly simple exponential dependence on $n$ given by $P_{e,min} = e^{-n \xi_{QCB}}$. The quantity $\xi_{QCB}$ is called the Quantum Chernoff Bound.  It is  a function of the density matrices only, with $\xi_{QCB} = -\log \big( \min_{0\le s \le 1} Tr (\rho^{s} \sigma^{1-s}) \big)$. The classical version of this problem was analyzed for the first time by H. Chernoff \cite{CHERNOFF}. It took 50 years to prove the quantum version of this problem. \\
\indent The Quantum Chernoff Bound has many interesting properties. Among these is monotonicity under CP maps, which makes it a valid distinguishability measure \cite{QCB}.  In addition, $\xi_{QCB}$ has an operational meaning arising from a statistical inference problem. Intuitively, we see that for a fixed probability of error $P_e$, the larger $\xi_{QCB}$ is the smaller the number of copies of $\rho$ and $\sigma$ we will need in order to distinguish them. \\
\indent In the present work, we apply the Quantum Chernoff Bound as a distinguishability measure to the manifold of Gibbs or thermal states associated with the anisotropic XY model in a transverse magnetic field. 
In the spirit of the information-theoretic and differential geometric approach advocated in \cite{F 3}, we will compare states corresponding to hamiltonians with slightly different values of the parameters $\{\lambda\}$  to derive a metric tensor for the parameter manifold itself.  This metric detects the second order quantum phase transitions of the XY model and shows the influence of finite temperature effects over the zero-temperature critical points. It is shown that  the different regions of the $\{T, \lambda\}$ phase diagram of the system above the T=0 critical points can be characterized by the different scaling behavior of the metric tensor with temperature. 
This paper parallels and extends the analysis reported in \cite{BURES} for the Bures metric. While the Quantum Chernoff Bound is associated with quantum state discrimination, the Bures metric is related with another natural probabilistic protocol: quantum estimation \cite{za-paris}. Our findings about the Quantum Chernoff Bound metric
show  that the same information about the phase diagram of an interacting many-body system can be obtained by using two independent  distinguishability measures.\\
\indent This paper is organized as follows. In Section II we introduce the many-body system we analyzed and obtain the corresponding metric tensor in Section III. In Section IV we carry out the thermal analysis of the metric elements. A global property of the metric is discussed in Section V, while conclusions and further research directions are outlined in Section VI.

\section{\label{sec:level1l}The XY Model}
We analyze the quantum phase transitions of the one-dimensional spin chain XY model in a transverse magnetic field given by the hamiltonian

\begin{equation}
H =  -\sum_{i=-\frac{N-1}{2}}^{i=\frac{N-1}{2}}\frac{1 + \gamma}{2} \sigma_i^x\sigma_{i+1}^x +\frac{1 - \gamma}{2} \sigma_i^y\sigma_{i+1}^y + \lambda \sigma_{i}^{z},\label{H}
\end{equation}

\noindent where the total number of spins $N$ is odd, $\lambda \in \Re$ is the transverse magnetic field along the z-axis, and $\gamma \in [-1,1]$ is the anisotropy parameter. 
For $\gamma=\pm1$ and $\gamma=0$ we obtain the Ising and XX models, respectively. 
This hamiltonian can be diagonalized using Jordan-Wigner,  Fourier and Bogoliubov transformations \cite{XY}. 
 The diagonalized hamiltonian is given by $H = \sum_{k=-\frac{N-1}{2}}^{k=\frac{N-1}{2}} \Lambda_{k}\hat{b}_{k}^{\dag}\hat{b}_{k}$,
with $\Lambda_{k}$ the quasi-particle energies given by $\Lambda_{k}=\sqrt{\epsilon_{k}^2+\Delta_{k}^2}$, with $\epsilon_k=\cos\big(\frac{2\pi k}{N}\big)-\lambda$ and $\Delta_k=\gamma\sin\big(\frac{2\pi k}{N}\big)$. 
One-particle excitations are created by the fermionic operators $\hat{b}^{\dag}_{k}=\cos(\frac{\theta_k}{2})\hat{d}_k^\dag+i\sin(\frac{\theta_k}{2})\hat{d}_{-k}$ acting on the ground state
\begin{equation} \label{gs}
|gs(\lambda,\gamma)\rangle=\bigotimes_{k=1}^{\frac{N-1}{2}}\big(\cos(\frac{\theta_k}{2})|00\rangle_{k, -k}
+i \sin(\frac{\theta_k}{2})|11\rangle_{k, -k}),
\end{equation}
with  $\hat{d}_{k}|00\rangle_{k,-k}=\hat{d}_{-k}|00\rangle_{k,-k}=\hat{b}_k|gs(\lambda,\gamma)\rangle=0$, and $\cos(\theta_k)=\epsilon_k/\Lambda_k$.
This model exhibits a quantum phase transition at two regions in the parameter space $\{ \gamma, \lambda \}$: at critical lines $\lambda=\pm1$, and $\gamma=0$ for $-1<\lambda<+1$. At those critical regions the system becomes gapless. In general  the gap $\Delta$ is given by: $\Delta = |1-|\lambda||$ if $|\lambda| > |1 - \gamma^2|$ (region A); $\Delta = |\gamma|  \sqrt{1 - \frac{\lambda^{2}}{1-\gamma^{2}}} $ if $|\lambda| < |1 - \gamma^2|$ (region B).

\section{\label{sec:level1}Quantum Chernoff Bound}
  The Quantum Chernoff Bound (QCB) is given by 
  \begin{equation}
  \xi_{QCB} = -\log \big( \min_{0\le s \le 1} Tr (\rho^{s} \sigma^{1-s}) \big),
  \end{equation} 
  where $P_{e,min} = e^{-n \xi_{QCB}}$ is the minimum probability of error in distinguishing two preparations $\rho^{\otimes n}$ and $\sigma^{\otimes n}$ in the limit $n \to \infty$ \cite{QCB}\cite{QCB2}.  Defining $Q(\rho,\sigma) = e^{-\xi_{{QCB}}}$, the quantity $1-Q$ serves as a distinguishability measure between states.  Notice that $1-Q(\rho, \rho) = 0$, and $1-Q(\rho, \sigma) = 1$ if $\rho$ and $\sigma$ are orthogonal.  Considering two nearby states $\rho$ and $\rho + d \rho$, this quantity induces a metric tensor on the manifold of density operators, with the line element given by
	\begin{equation}
	ds^2 = \frac{1}{2} \sum_{i j} \frac{|\bra{i} d\rho \ket{j}|^2}{(\sqrt{p_{i}} + \sqrt{p_{j}})^2},
	\end{equation}
	where $\rho = \sum_{i} p_{i} \ket{i} \bra{i}$ is a spectral decomposition of the density operator.  Since we are interested in the finite temperature XY model, we consider thermal states of the form $\rho = \frac{e^{-\beta H}}{Z}$.  Taking the spectral decomposition of $\rho$ over eigenstates of the hamiltonian, we can split the metric into two parts so that $ds^2 = ds_{c}^2 + ds_{nc}^2$, where
	\begin{eqnarray}
	ds_{c}^2 & = & \frac{1}{8} \sum_{i} \frac{(d p_{i})^2}{p_{i}} \label{dsc}\\
	ds_{nc}^2 & = & \frac{1}{2} \sum_{i \neq j} \frac{|\braket{i}{dj}|^2 (p_{i}-p_{j})^2}{(\sqrt{p_{i}}+\sqrt{p_{j}})^2}.
	\end{eqnarray}
	\indent We call the first sum the classical part of the metric since it only depends on the Boltzmann weights of the density operator.  We label the second sum the non-classical or quantum part, because it explicitly depends on the states.\\
\indent In passing we would like to notice that $ds_{nc}^2$ does not, strictly speaking, define a metric in the parameter space. Indeed one can have curves 
$t\rightarrow\rho(t)$  of density matrices
where all the relevant $|j\rangle$ are fixed, e.g. $\rho(t)=\sum_i p_i(t)|i\rangle\langle i|.$ In other terms, 
the quadratic form defined at each point of the parameter manifold is just positive-semidefinite rather than positive definite and the metric matrix can have zero eigenvalues.
This remark has to be kept in mind when one considers the zero-temperature limit $\beta\rightarrow\infty$ where, as we will see, $ds\rightarrow ds_{nc}$. Moreover, in this limit $g_{nc}$ is nothing but the Fubini-Study metric over the projective state space 
(the same is obtained by starting from the Bures metric). Once this latter is pulled back to the parameter space null eigenvectors can appear.
These vectors correspond to directions along which changing the parameters results in indistinguishable states.\\
	\indent To evaluate the first sum (\ref{dsc}), note that
	\begin{eqnarray*}
	\partial_{\beta} p_{i} = -(E_{i} - \langle E \rangle) \frac{e^{-\beta E_{i}}}{Z} \\
	\partial_{\gamma} p_{i} = -\beta (\partial_{\gamma} E_{i} - \langle \partial_{\gamma} E \rangle ) \frac{e^{-\beta E_{i}}}{Z} \\
	\partial_{\lambda} p_{i} = -\beta (\partial_{\lambda} E_{i} - \langle \partial_{\lambda} E \rangle ) \frac{e^{-\beta E_{i}}}{Z},
	\end{eqnarray*}
	where $E_i=\sum_k n_k \Lambda_k,\,(n_k\in\{0,1\}).$ \\
	\indent Summing over states and incorporating the fermion statistics $\langle n_{\mu} \rangle = (1+e^{\beta \Lambda_{\nu}})^{-1}$
	and using the free-fermion property $\langle n_{\mu} n_{\nu} \rangle - \langle n_{\mu} \rangle \langle n_{\nu} \rangle =\delta_{\mu \nu} \langle n_{\mu} \rangle (1 - \langle n_{\mu} \rangle),$
	 we get the following six components for the 3x3 symmetric metric tensor defined by $ds_{c}^2 = g_{\mu \nu}^{c} dx^{\mu} dx^{\nu}$, $x^{\mu} \in \{ \beta, \gamma, \lambda \}$
	\begin{eqnarray}
	g_{\beta \beta}^{c} & = & \frac{1}{16} \sum_{k} \frac{1}{\cosh(\beta \Lambda_{k}) + 1} \Lambda_{k}^2 \\
	g_{\beta \gamma}^{c} & = & \frac{\beta}{16 \gamma} \sum_{k} \frac{1}{\cosh(\beta \Lambda_{k}) + 1} \Delta_{k}^2\\
	g_{\beta \lambda}^{c} & = & \frac{-\beta}{16} \sum_{k} \frac{1}{\cosh(\beta \Lambda_{k})+1} \epsilon_{k} \\
	g_{\gamma \gamma}^{c} & = & \frac{\beta^2}{16 \gamma^2} \sum_{k} \frac{1}{\cosh(\beta \Lambda_{k} + 1)} \frac{\Delta_{k}^4}{\Lambda_{k}^2}\\
	g_{\gamma \lambda}^{c} & = & \frac{-\beta^2}{16 \gamma} \sum_{k} \frac{1}{\cosh(\beta \Lambda_{k}) + 1} \frac{\epsilon_{k} \Delta_{k}^2}{\Lambda_{k}^2}\\
	g_{\lambda \lambda}^{c} & = & \frac{\beta^2}{16} \sum_{k} \frac{1}{\cosh(\beta \Lambda_{k}) + 1} \frac{\epsilon_{k}^2}{\Lambda_{k}^2}.
	\end{eqnarray}
	\indent In order to find the metric $g_{\mu \nu}^{nc}$ corresponding to $ds_{nc}^2$ we first use the expression for the eigenstates (\ref{gs}) to obtain
	\begin{equation}
	ds_{nc}^2 = \frac{1}{4} \sum_{k} \frac{\cosh (\beta \Lambda_{k})-1}{\cosh (\beta \Lambda_{k})+1} d \theta_{k}^2.
	\end{equation}
	\indent Differentiating $\theta_{k}$ along each of our three parameters gives the metric $g_{\mu \nu}^{nc}$.  This part of the metric has only three nonzero components, namely
	\begin{eqnarray}
	g_{\gamma \gamma}^{nc} & = & \frac{1}{4 \gamma^2} \sum_{k} \Big(\frac{\cosh(\beta \Lambda_{k})-1}{\cosh(\beta \Lambda_{k})+1} \Big) \frac{\epsilon_{k}^2 \Delta_{k}^2}{\Lambda_{k}^4} \\
	g_{\gamma \lambda}^{nc} & = & \frac{1}{4 \gamma} \sum_{k} \Big(\frac{\cosh(\beta \Lambda_{k})-1}{\cosh(\beta \Lambda_{k})+1} \Big) \frac{\epsilon_{k} \Delta_{k}^2}{\Lambda_{k}^4} \\
	g_{\lambda \lambda}^{nc} & = & \frac{1}{4} \sum_{k} \Big(\frac{\cosh(\beta \Lambda_{k})-1}{\cosh(\beta \Lambda_{k})+1} \Big) \frac{\Delta_{k}^2}{\Lambda_{k}^4} .
	\end{eqnarray}
\indent Notice that $g_{\lambda\lambda}^{nc}$ vanishes along the line $\gamma=0.$

\section{\label{sec4} Thermal analysis of the metric elements}
We now proceed to analyze the scaling behavior of the metric elements with temperature, for two characteristic regions in the parameter space $\{\beta$, $\gamma$, $\lambda\}$: quantum-critical and quasi-classical. For the quantum-critical case we analyze the scaling at the critical region.  For the quasi-classical case we carry out the scaling analysis away from the critical region, for temperatures small enough so that the system looks effectively gapped, i.e.,  $\beta\Delta\gg 1$.
We carry out the analysis in the thermodynamic limit, in which $\frac{2\pi k}{N}\rightarrow k$, $\sum_k\rightarrow \frac{1}{2\pi}\int_{-\pi}^{\pi}dk$ after rescaling $g\rightarrow g/N$. \\

\subsection{Quasi-Classical Region}
We now find the scaling behavior with temperature for the metric element $g_{\gamma\gamma}^{nc}$, in the limit $\beta\Delta\gg 1$. The scaling of all other metric elements can be obtained following procedures similar to those illustrated below. \\
\indent Since the system is gapped in this region, we have $\beta\Lambda_{k}>\beta\Delta\gg1$. This permits us to make the following approximation in the integral expression for $g_{\gamma\gamma}^{nc}$: $ \Big(\frac{\cosh(\beta \Lambda_{k})-1}{\cosh(\beta \Lambda_{k})+1} \Big) \cong \frac{e^{\beta\Lambda_k}-2}{e^{\beta\Lambda_k}+2} \cong 1-4e^{-\beta\Lambda_k}$. Then, $g_{\gamma\gamma}^{nc}$ can be approximated by

\begin{equation}
g_{\gamma\gamma}^{nc} \cong g_{\gamma\gamma}^{nc}(T=0) - \frac{1}{2\pi}\int_{-\pi}^{\pi}e^{-\beta\Lambda_{k}}\Big(\frac{\epsilon_k\sin(k)}{\Lambda_k} \Big)^{2}.\nonumber \\
\end{equation}
\indent The exponential $e^{-\beta\Lambda_{k}}$ behaves like a sharp peak in $k$ centered around the absolute minima of $\Lambda_k$ and can be approximated further by expanding $\Lambda_k$ to second order
in $k$. The minima of $\Lambda_k$ occur at $k=0$ for region A, $\lambda>0$; at $k = \pm\arccos(\frac{\lambda}{1-\gamma^2})$ for region B; and at $k=\pm\pi$ for region A, $\lambda<0$.\\
\indent Due to the peak of $e^{-\beta\Lambda_{k}}$, the rest of the integrand can be approximated by its value at the minimum of $\Lambda_k$ in region B, or expanded until second order in $k$ in region A. 
Taking the limits of integration to be from $-\infty$ to $\infty$, this approximation results in Gaussian integrals, with the result  given by
\begin{equation}\label{nc}
g_{\gamma\gamma}^{nc}(\beta\Delta\gg 1)= g_{\gamma\gamma}^{nc}(T=0)-f(\lambda,\gamma)T^{\alpha}e^{- \Delta/T},
\end{equation}

\noindent where the exponent $\alpha$ of the temperature is equal to $3/2$ for region A, and $1/2$ for region B. Each metric element will have a different $f(\lambda,\gamma)$.
Similarly, the classical terms of the metric elements can be approximated by
\begin{equation}\label{c}
g^{c}(\beta\Delta\gg 1)=h(\lambda,\gamma)T^{\alpha}e^{- \Delta/T}.
\end{equation}

\indent In Table  \ref{tab:table1} we summarize the scaling in temperature for all the metric elements in the quasi-classical region.
\begin{table}[h]
\caption{Temperature exponent $\alpha$, for classical and nonclassical terms of the metric elements in the quasi-classical region. The nonclassical terms follow a behavior given by (\ref{nc}), while the classical terms behave like (\ref{c}). } \label{tab:table1}
\begin{ruledtabular}
\begin{tabular}{cccccccccc}
$\textrm{Region}$ &$g_{\beta\beta}^{c}$ &$g_{\beta\lambda}^{c}$ & $g_{\beta\gamma}^{c}$ & $g_{\lambda\lambda}^{c}$ & $g_{\lambda\gamma}^{c}$ & $g_{\gamma\gamma}^{c}$ & $g_{\lambda\lambda}^{nc}$ & $g_{\lambda\gamma}^{nc}$ & $g_{\gamma\gamma}^{nc}$\\
\hline
$\textrm{A}$ & 1/2 & -1/2 & 1/2 & -3/2 & -1/2 & 1/2 & 3/2 & 3/2 & 3/2 \\
$\textrm{B}$ & 1/2 & -1/2 & -1/2 & -3/2 & -3/2 & -3/2 & 1/2 & 1/2 & 1/2 \\
\end{tabular}
\end{ruledtabular}
\end{table}

\subsection{Quantum-Critical Region}
	Consider now taking $T \to 0$ at the values of $\gamma$ and $\lambda$ for which $\Delta = 0$.  Since in this limit $\beta \to \infty$, all classical metric elements vanish due to the factor of $(\cosh( \beta \Lambda_{k} ) + 1)^{-1}$ in front.  We are left only to analyze the nonclassical part of the metric, namely $g_{\gamma \gamma}^{nc}$, $g_{\gamma \lambda}^{nc}$, and $g_{\lambda \lambda}^{nc}$.  There are three cases to consider: i) $\gamma = 0$ and $\lambda = \pm 1$, ii) $\gamma = 0$ and $-1 < \lambda < 1$, and iii) $\gamma \ne 0$ and $\lambda = \pm 1$.\\
\indent Since $g_{\gamma \lambda}^{nc}$ and $g_{\lambda \lambda}^{nc}$ are multiplied by an overall factor of $\gamma$, both vanish for cases i) and ii).\\
\indent The dispersion $\Lambda_{k}$ is an even function of $k$, so we can restrict ourselves to the interval $k \in [0,\pi]$.  For cases ii) and iii) $\Lambda_{k}$ is linear in $k$ about its root, whereas for the case i) $\Lambda_{k}$ is quadratic about its root.\\
\indent We now show how to calculate the scaling in temperature of the metric element $g_{\gamma \gamma}^{nc}$ for case i).  A similar procedure applies to $g_{\lambda \lambda}^{nc}$ and $g_{\gamma \lambda}^{nc}$ for the other cases.\\
\indent The goal is to bound the metric element above and below by functions that have the same scaling behavior in $\beta$.  This will ensure that the metric itself must scale with the same exponent.\\
\indent Since the dispersion $\Lambda_{k}$ is quadratic around the root $k = 0$, we can approximate it as $\Lambda_{k} \thicksim \frac{k^2}{2}$.  Define the piecewise function
\begin{equation}\label{f}
f(\beta, k) = \left \{
\begin{array}{ll}
\frac{\beta^2 k^4}{16} & \textrm{for $0 \le k \le \frac{2}{\sqrt{\beta}}$}\\
1 & \textrm{for $\frac{2}{\sqrt{\beta}} \le k \le \pi$},\\
\end{array} \right.
\end{equation}
which for all $\beta$ and $k$ satisfies
\begin{equation}\label{bounding}
\frac{f(\beta, k)}{2} < \frac{\cosh(\frac{\beta k^2}{2})-1}{\cosh(\frac{\beta k^2}{2})+1} < f(\beta, k).
\end{equation}
\indent We can now split the integral into two parts
\begin{eqnarray}
g_{\gamma \gamma}^{nc} &=& \frac{1}{4 \pi} \int_{0}^{\pi} dk \frac{\cosh ( \beta \Lambda_{k}) - 1}{\cosh ( \beta \Lambda_{k}) + 1} \frac{\epsilon_{k}^2 \sin^2(k)}{\Lambda_{k}^4} \nonumber \\
& \simeq & \frac{1}{4 \pi} \int_{0}^{\frac{2}{\sqrt{\beta}}} dk \frac{\cosh ( \frac{\beta k^2}{2}) - 1}{\cosh ( \frac{\beta k^2}{2}) + 1} \frac{4}{k^2} \label{ggg1} \\
& + & \frac{1}{4 \pi} \int_{\frac{2}{\sqrt{\beta}}}^{\pi} dk \frac{\cosh ( \beta \Lambda_{k}) - 1}{\cosh ( \beta \Lambda_{k}) + 1} \frac{\epsilon_{k}^2 \sin^2{k}}{\Lambda_{k}^4}, \label{ggg2}
\end{eqnarray}
where we have taken $\Lambda_{k} \thicksim \frac{k^2}{2}$ and $\sin(k) \thicksim k$ for the integral (\ref{ggg1}).  Note that this is a good approximation for $\beta \to \infty$, since the upper integration limit becomes arbitrarily close to $0$. \\
\indent We can now bound the first integral above and below, using the function (\ref{f})
\begin{eqnarray*}
\frac{1}{4 \pi} \int_{0}^{\frac{2}{\sqrt{\beta}}} dk \frac{f(\beta, k)}{2} \frac{4}{k^2} & \le & \frac{1}{4 \pi} \int_{0}^{\frac{2}{\sqrt{\beta}}} dk \frac{\cosh (  \frac{\beta k^2}{2} ) - 1}{\cosh (  \frac{\beta k^2}{2}) + 1} \frac{4}{k^2} \\
& \le & \frac{1}{4 \pi} \int_{0}^{\frac{2}{\sqrt{\beta}}} dk f(\beta, k) \frac{4}{k^2}.
\end{eqnarray*}
\indent These bounding integrals scale as $\beta^{\frac{1}{2}}$.  Therefore the first integral (\ref{ggg1}) must also scale as $\beta^{\frac{1}{2}}$ for $\beta \to \infty$. \\
\indent Over the entire interval $[0, \pi]$ we can bound $\Lambda_{k}^{-1}$ from above by $\frac{6}{k^2}$, upper bound $\sin^2(k)$ by $k^2$, and replace the ratio of hyperbolic cosines in (\ref{ggg2}) by $f(\beta, k)$.  This will bound integral (\ref{ggg2}) from above, as
\begin{eqnarray*}
\frac{1}{4 \pi} \int_{\frac{2}{\sqrt{\beta}}}^{\pi} dk \frac{\cosh ( \beta \Lambda_{k}) - 1}{\cosh ( \beta \Lambda_{k}) + 1} \frac{\epsilon_{k}^2 \sin^2{k}}{\Lambda_{k}^4} & \le & \frac{1}{4 \pi} \int_{\frac{2}{\sqrt{\beta}}}^{\pi} dk \frac{36}{k^2} \\
& \thicksim & \beta^{\frac{1}{2}}.
\end{eqnarray*}
Therefore, since the second integral (\ref{ggg2}) scales as a power of ${\frac{1}{2}}$ or lower in $\beta$, $g_{\gamma \gamma}^{nc}$ must scale as $\beta^{\frac{1}{2}}$ to highest order.  Table \ref{tab:table3} lists the scaling for all three non-classical metric elements.

\begin{table}[h]
\caption{ \label{tab:table3}
Scaling behavior with temperature at criticality, for the metric elements
$g_{\lambda\lambda}^{nc}$, $g_{\gamma\lambda}^{nc}$ and $g_{\gamma\gamma}^{nc}$. Note that $g_{\lambda\lambda}^{nc}$ and $g_{\gamma\lambda}^{nc}$ are exactly 0 for $\gamma=0$. }
\begin{ruledtabular}
\begin{tabular}{cccc}
 &$\lambda=\pm 1, \gamma=0$&$\lambda = \pm 1, \gamma\ne 0$ & $\lambda \in (-1,1), \gamma=0$\\
\hline
$g_{\lambda\lambda}^{nc}$ & 0 & $T^{-1}$ & 0 \\
$g_{\gamma\lambda}^{nc}$ & 0 & const$+ O(T)$ & 0 \\
$g_{\gamma\gamma}^{nc}$ & $T^{-1/2}$& const$+O(T^3)$ & $T^{-1}$ \\
\end{tabular}
\end{ruledtabular}  
\end{table}
The behavior of the nonclassical metric elements in the quantum-critical region can be inferred from
dimensional scaling analysis as was done in \cite{BURES}, for zero temperature. Though carried out for the Bures metric, it
applies also for the Quantum Chernoff Bound metric. From \cite{BURES} we see that the scaling dimension of the nonclassical metric elements $g_{\lambda\lambda}^{nc}, g_{\gamma\lambda}^{nc}$, and $g_{\gamma\gamma}^{nc}$ is given by $\Delta_{nc}=\Delta_\mu+\Delta_\nu-2z-d$, where $\Delta_\mu, \Delta_\nu$ are the scaling dimensions of the operators that couple to $\mu, \nu \in \{\lambda,\gamma\}$  and produce the quantum phase transitions, $z$ is the dynamical exponent, and $d$ the spatial dimensionality. The finite temperature in the quantum system transforms into a finite additional length dimension in the classical system so that
\begin{equation}
g^{nc}\sim T^{\Delta_{nc}/z}.
\end{equation}
The findings summarized in Table \ref{tab:table3}
are consistent with this relation, assuming that: $\Delta_\gamma=2,\, \Delta_\lambda\ge 3$  (1st column, $z=2$); $\Delta_\gamma=3,\, \Delta_\lambda=1$ (2nd column, $z=1$);
$\Delta_\gamma=1,\, \Delta_\lambda\ge 2$ (3rd column, $z=1).$ The condition $\Delta_x < d+z$ ($d=1$) signals the relevance, in the renormalization group sense \cite{gold},
of the operator weighted by the coupling constant $x$ $(x=\gamma,\,\lambda).$ \\
\indent Intuitively, shifting some parameter towards a critical value should result in a significant change in the state of our system.  That is, the direction in parameter space corresponding to maximal distinguishability should be towards or away from the critical regions.  Indeed we see that the diverging metric element in the gapless quantum critical case corresponds to exactly this direction of maximal distinguishability.  As an example, for $\lambda = \pm 1$ and $\gamma \ne 0$ only $g_{\lambda \lambda}$ diverges.  Changing $\lambda$ and keeping $\gamma$ constant here corresponds to the direction of maximal distinguishability, which is along the $g_{\lambda \lambda}$ contribution to the line element. In the Ising model limit, where $\gamma = 1$, our results for the metric scaling agrees exactly with that of \cite{BURES}, where the Bures metric was used.

\section{Maximum eigenvalue}

In this brief section we consider
 the maximum eigenvalue of the metric $g$ for each point in parameter space.  Since $g$ is a symmetric matrix, this corresponds to the matrix norm $\|g\|
=\max_{\|{\bf{v}}\|=1} |\langle {\bf{v}}, g{\bf{v}}\rangle|$. 
The associated eigenvectors define the field of maximum distinguishability directions. An analogous analysis for the Bures metric has been presented in \cite{BURES}.
From the  inequalities $\max_{i,j} |g_{ij}|\le \|g\|\le 9 \max_{i,j} |g_{ij}|$ one immediately sees 
that the maximal eigenvalue of $g$ diverges iff at least one of the matrix elements of $g$ diverges. In other words, $\|g\|$ encodes global information about the metric
that one can extract without analyzing each of the matrix elements separately.  

\begin{figure}[htp]
\centering
\includegraphics[totalheight=0.3\textheight]{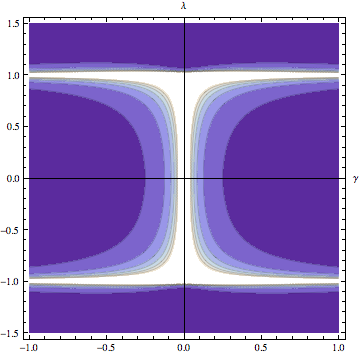}
\caption{Contour plot of the maximum eigenvalue of $g$ for $-1.5 < \lambda < 1.5$, $-1 < \gamma < 1$, and $T = 10^{-2}$.  The color scale goes from 0 (purple), to 3 (white). The critical regions can be clearly identified.}\label{fig:contour}
\end{figure}

Shown in Figure \ref{fig:contour} is a contour plot of the maximum eigenvalues of $g$. The critical regions are clearly revealed as those values of the parameters for which some metric elements diverge for small temperature.

\section{Conclusion}
\indent 
In this paper we have employed the metric tensor induced on the parameter space by the Quantum Chernoff Bound to study the finite-temperature phase diagram of the anisotropic XY spin chain.
We have shown that the temperature scaling of the metric elements has a different form in the quasi-classical and quantum-critical cases.  In the quasi-classical case, where the temperature
$T$ is much smaller than the energy gap $\Delta,$ the metric elements vanish exponentially in $1/T;$ in the quantum-critical regime $\Delta/T\ll 1,  T \to 0$, the behavior is a power law in $T$.  Moreover, in the quasi-classical case the specific behavior in temperature depends on the functional form of the gap $\Delta$, and in the quantum critical case the nonclassical metric element which diverges in the limit of $T \to 0$ is the one corresponding to the relevant parameter driving the QPT. 
We would like to notice that,
in view of the operational meaning of the Quantum Chernoff Bound metric in quantum state discrimination tasks,
 these results suggest that quantum criticality
may provide a resource for quantum state discrimination protocols.
This connection has been elaborated in \cite{za-paris} for the related quantum estimation problem.
Another problem that is worthy of investigation and has not been touched in this paper is the role of the curvature tensor associated to the metric $g.$
The preliminary findings for the same quantity in the pure state case \cite{F 3} and in the Bures metric case \cite{BURES} suggest that the curvature might provide additional information about the nature of the 
critical lines and location of crossover regions.

The analysis reported in this paper for the $XY$ chain is quite close in spirit and technicalities to the one for the Bures metric in the Quantum Ising model case  \cite{BURES}.
In both cases the important physical features of the low-temperature phase diagram of the systems, e.g., quantum-critical to quasi-classical crossovers, 
 can be identified by resorting to a metric with a statistical distinguishability meaning. 
This proves that the information-theoretic metrical approach to quantum criticality \cite{F 3} 
is not bound to the use of a special single metric. Different choices of distinguishability measure may provide essentially the same physical information.
\section{Acknowledgements}
We would like to thank for fruitful discussions M. Cozzini, P. Giorda and L. Campos Venuti.
D.F.A. and N.T.J. are grateful to the Quantum Unit of the ISI Foundation for its hospitality and for providing a stimulating scientific environment.


\begin{thebibliography}{99}
\bibitem{QPT} S. Sachdev, \textit{Quantum Phase Transitions}, Cambridge University Press, Cambridge, England, 1999. M. Vojta, Rep. Prog. Phys. { \bf{66}}, (2003), 2069-2110.
\bibitem{EMB 1} A. Osterloh, L. Amico, G. Falci, and R. Fazio, Nature {\bf{416}}, 608 (2002). 
\bibitem{EMB 2} T. J. Osborne and M. A. Nielsen, Phys. Rev. A {\bf{66}}, 032110 (2002); Quantum Inf. Process. {\bf{1}}, 45 (2002).
\bibitem{EMB 3} G.~Vidal, J.I.~Latorre, E.~Rico, and A.~Kitaev,  Phys. Rev. Lett. {\bf 90}, 227902 (2003). 
\bibitem{EMB 4} L. -A. Wu, M. S. Sarandy and D. A. Lidar, Phys Rev Lett. {\bf{93}}, 250404 (2004).
\bibitem{EMB 5} T.R. de Oliveira, G. Rigolin, M. C. de Oliveira, and E. Miranda, Phys. Rev. Lett. {\bf{97}}, 170401 (2006).
\bibitem{EMB 6} L. Amico, R. Fazio, A. Osterloh and J. V. Vedral, quant-ph/0703044.
\bibitem{F 1} P.~Zanardi and N.~Paunkovic, Phys. Rev. E {\bf 74}, 031123 (2006).
\bibitem{F 2} Y.~Chen, P.~Zanardi, Z.D.~Wang, F.C.~Zhang, New J. Phys. {\bf 8}, 97 (2006). 
\bibitem{F 3}P. Zanardi, P. Giorda and M. Cozzini, Phys. Rev. Lett. {\bf{99}}, 100603 (2007).
\bibitem{F3a} L.  Campos Venuti, P. Zanardi, Phys. Rev. Lett. {\bf{99}}, 095701 (2007)
\bibitem{F 4} P. Zanardi, M. Cozzini and P. Giorda, J.~Stat.~Mech.~L02002 (2007).
\bibitem{F 5} M. Cozzini, P. Giorda and P. Zanardi, Phys.~Rev.~B \textbf{75}, 014439 (2007).
\bibitem{F 6}  M.~Cozzini, R.~Ionicioiu and P.~Zanardi, Phys. Rev. B {\bf{76}}, 104420 (2007).
\bibitem{F 7} P.~Zanardi, H.-T.~Quan and X.-G.~Wang and C.-P.~Sun, Phys. Rev. A {\bf{75}}, 032109 (2007).
\bibitem{F 8} P. Buonsante and A. Vezzani, Phys.~Rev.~Lett.~\textbf{98}, 110601 (2007).
\bibitem{F 9} M.-F. Yang, arXiv:0707.4574;  H.-Q. Zhou and J. P. Barjaktarevic, cond-mat/0701608.
\bibitem{F 10} H.-Q. Zhou, J.-H. Zhao and B.Li, arXiv:0704.2940; H.-Q. Zhou, arXiv:0704.2945.
\bibitem{F 11} Y.-C. Tzeng, M. -F. Yang, arXiv:0709.1518.
\bibitem{SCALING}L. Campos Venuti and P. Zanardi, Phys. Rev. Lett. {\bf{99}}, 095701 (2007)
\bibitem{BURES} P. Zanardi, L. Campos Venuti, P. Giorda, to appear in Phys. Rev. A, arXiv:0707.2772v2
\bibitem{za-paris} P. Zanardi, M.G.A Paris, arXiv:0708.1089
\bibitem{gold} N. Goldenfeld,  \textit{Lectures on phase transitions and the renormalization group}, Westview Press, Boulder, 1992
\bibitem{QCB} K.M.R. Audenaert, J. Calsamiglia, et. al., Phys. Rev. Lett. { \bf{98} }, 160501 (2007).

\bibitem{QCB2} M. Nussbaum and A. Szko\l a, quantum-ph/0607216.
\bibitem{CHERNOFF} H. Chernoff, Ann. Math. Stat. {\bf{23}}, 493, (1952)

\bibitem{XY} J. I. Latorre, E. Rico, and G. Vidal, Quant. Inf. Comput. { \bf{4} } (2004) 48-92

\end{thebibliography}
\end{document}